\documentclass[preprint, a4paper, 10pt]{elsarticle}

\usepackage{amsmath}

\usepackage{multicol}

\usepackage{setspace}

\usepackage[paper=a4paper,left=15mm,right=15mm,top=25mm,bottom=20mm]{geometry}

\usepackage{color} 
\PassOptionsToPackage{dvips}{graphicx}
\usepackage[dvips]{graphicx}
\usepackage{lineno,hyperref}
\modulolinenumbers[5]
\journal{Surface Science}
\makeatletter
\def\@author#1{\g@addto@macro\elsauthors{\normalsize%
    \def\baselinestretch{1}%
    \upshape\authorsep#1\unskip\textsuperscript{%
      \ifx\@fnmark\@empty\else\unskip\sep\@fnmark\let\sep=,\fi
      \ifx\@corref\@empty\else\unskip\sep\@corref\let\sep=,\fi
      }%
    \def\authorsep{\unskip,\space}%
    \global\let\@fnmark\@empty
    \global\let\@corref\@empty  %% Added
    \global\let\sep\@empty}%
    \@eadauthor={#1}
}
\makeatother

\begin{document}

%\begin{titlepage}

\begin{frontmatter}

\title{STM study of the preparation of clean Ta(110) and the subsequent growth of two-dimensional Fe islands  %%\tnoteref{mytitlenote}
}
%%\tnotetext[mytitlenote]{Fully documented templates are available in the elsarticle package on %%\href{http://www.ctan.org/tex-archive/macros/latex/contrib/elsarticle}{CTAN}.}

\author{T.~Eelbo\corref{cor1}}
%\ead{teelbo@physnet.uni-hamburg.de}
\cortext[cor1]{Corresponding author: teelbo@physnet.uni-hamburg.de}
\author{V.~I.~Zdravkov\fnref{VZ1,VZ2}}
\fntext[VZ1]{current address: Experimental Physics II, Institute of Physics, University of Augsburg, Universit\"{a}tsstra{\ss}e 1, D-86135 Augsburg, Germany}
\fntext[VZ2]{permanent address: Institute of Electronic Engineering and Nanotechnologies ''D. Ghitu'', Academy of Sciences of Moldova, Academiei str., 3/3, MD-2028, Chisinau, Republica Moldova}
\author{R.~Wiesendanger}
\address{Department of Physics, Scanning Probe Methods Group, University of Hamburg, Jungiusstra{\ss}e 11a, D-20355 Hamburg}

%%\fntext[myfootnote]{Since 1880.}

%%\author[mymainaddress,mysecondaryaddress]{Elsevier Inc}
%%\ead[url]{www.elsevier.com}

\date{\today}

\doublespacing

\begin{abstract}

This report deals with the preparation of a clean Ta(110) surface, investigated by means of scanning tunneling microscopy/spectroscopy as well as by low-energy electron diffraction and Auger electron spectroscopy. The surface initially exhibits a surface reconstruction induced by oxygen contamination. This reconstruction can be removed by annealing at high temperatures under ultrahigh vacuum conditions. The reconstruction-free surface reveals a surface resonance at a bias voltage of about -500 mV. The stages of the transformation are presented and discussed. In a next step, Fe islands were grown on top of Ta(110) and investigated subsequently. An intermixing regime was identified for annealing temperatures of (550 -- 590)~K.

\end{abstract}

\begin{keyword}
\texttt{Scanning tunneling microscopy, Scanning tunneling spectroscopy, Tantalum, Surface segregation, Iron islands, growth study}
\end{keyword}

\end{frontmatter}
%\begin{multicols}{2}

\section{Introduction}

\doublespacing

Within the past years, the observation of novel phenomena related to finite size effects on surfaces, ultra-thin films and nanoparticles involving elementary superconductors (SCs) excite intense scientific interest, e.g. the parity effect in SC~\cite{Janko1994} or the shell effect of SC nanoparticles~\cite{Bose2010, Brihuega2011}. Local probe techniques are indispensable tools to investigate these effects on the atomic scale. Notably, scanning tunneling microscopy/spectroscopy (STM/STS) has a great potential to study such systems due to its lateral, energy, and spin resolution capabilities~\cite{Binnig1987, Wiesendanger1990}. Indeed, greatly improved understanding was achieved in the underlying physics of topological defects in SC~\cite{Suderow2014, Roditchev2015}, of the origination of unconventional superconductivity~\cite{Kalcheim2012, Ruby2015} and of novel finite-size phenomena~\cite{Qin2009, Araujo2011, Brun2014, Bose2014}. This enabled, for instance, artificial control of low dimensional, proximity induced superconductivity~\cite{Linder2015} and exotic non-Abelian fermions~\cite{Nadj2014}.

The main goal of this report is the investigation of the properties and phenomena arising from the interaction of a bulk elementary SC and magnetic nanostructures grown on top. Ta was chosen since it is one of the few elementary SCs with a transition temperature above that of the liquid $^4$He phase. Furthermore, Ta can establish strong spin-orbit coupling at its surface~\cite{Wortelen2015}. In a first step, the aim was preparing a clean Ta(110) surface. In a second step, sub-monolayer coverages of Fe were grown on top and the resulting hybrid system was studied by STM/STS.

Ta has a bcc structure, with a lattice constant of $a = 330.1$~pm. The (110) surface exhibits a two-fold symmetric lattice with nearest neighbor distances of $285.9$~pm along [1$\bar{1}$1]- and [1$\bar{1}\bar{1}$]-directions, which comprise an angle $\alpha \approx 70^\circ$. Ta is known to be a very strong getter material and exhibits high bulk solubility and diffusivity. Previous investigations revealed that the main contamination of Ta is oxygen. The oxygen is dissolved in the bulk, but represents the main source of contamination at Ta surfaces as well~\cite{Sewell1972, Huger2005}. The Ta-O phase diagram describes only one stable oxide, namely Ta$_{2}$O$_{5}$, but under thin film conditions or due to the presence of other types of impurities, there may be further metastable oxides~\cite{Garg1996}. 

\section{Experimental setup}

The experiments were performed in an ultrahigh vacuum system with a base pressure in the low 10$^{-11}$~mbar range. Auger electron spectroscopy (AES) and low energy electron diffraction (LEED) were carried out at room temperature while STM/STS investigations were conducted either at room temperature or at 37~K. The samples were prepared \textit{in-situ}, initially by combined Ar$^+$-ion sputtering and subsequent annealing cycles, later on by heating using an electron-beam heater only. The temperature was regularly checked during electron bombardment using an optical pyrometer. During the entire preparation procedure the oxygen partial pressure was constrained to the low 10$^{-9}$~mbar range to reduce adsorption of oxygen from the residual gas. To gain information about the local density of states, electronic conductance spectra were acquired by means of a lock-in technique using a modulation voltage $U_{mod} = 20$~mV and a frequency $f = 3.333$~kHz.

\section {Results and discussion}

\subsection{Preparation of the Ta(110) surface}

\begin{figure*}[ht!]
\centering
\includegraphics[width=0.75\textwidth]{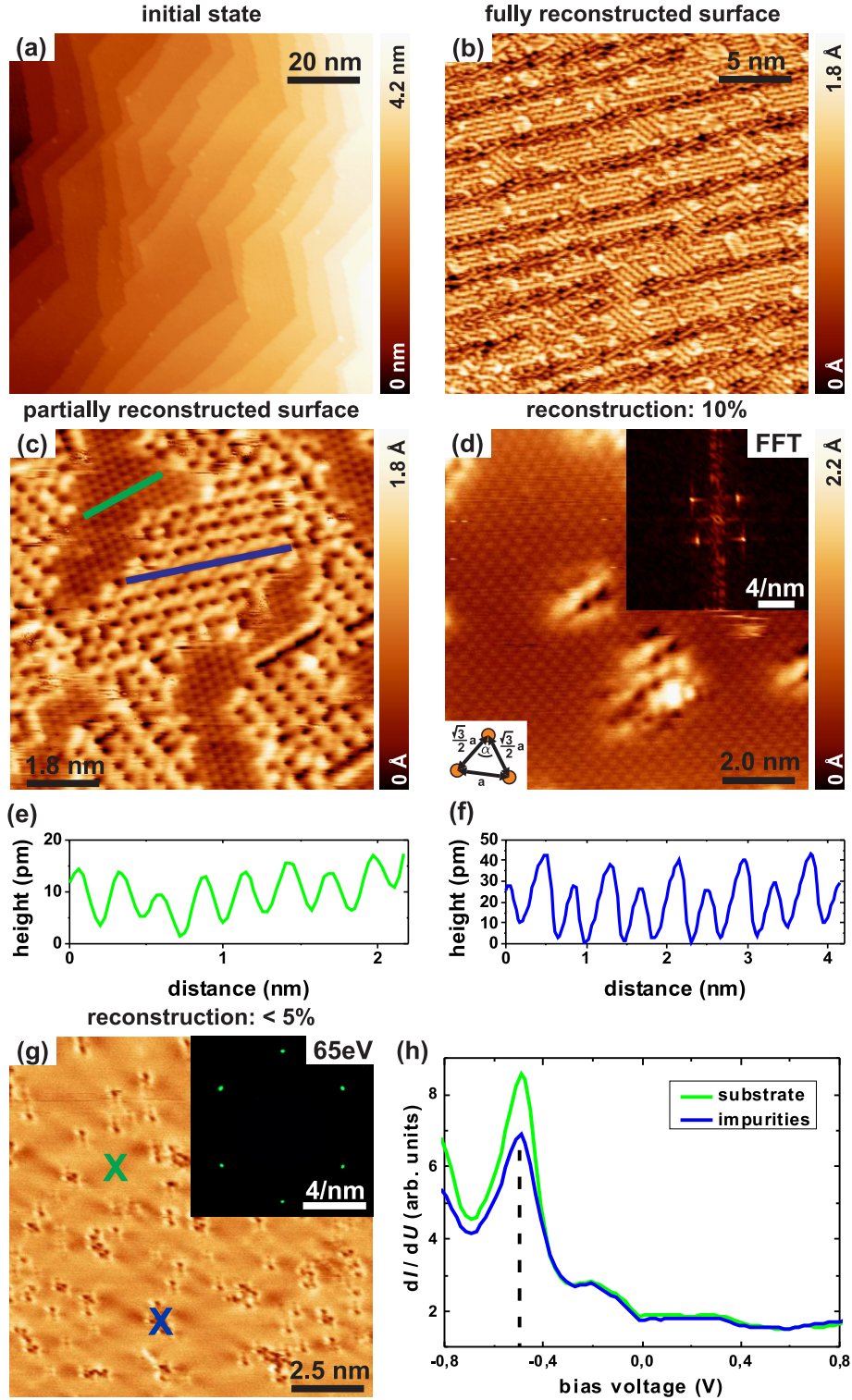}
\caption{(a)-(d), (g): STM topographies of Ta(110) acquired at room temperature showing the surface at different levels of surface purity. (a) shows the Ta(110) surface in its initial state. (b) exhibits an entirely reconstructed surface after a few cycles of annealing. (c) shows the varying relative orientation of the reconstructed domains and the atomic lattice in clean areas. (d) shows the atomic lattice of the bare Ta(110) surface. The zoom-in in (d) shows the 2D-FFT of the surface and reveals pronounced spots caused by the atomic lattice. (e) and (f) show line profiles of the two different lattices as indicated in (c). (g) is a representative topography of the best surface quality achieved. The zoom-in in (g) shows the corresponding LEED pattern. (h) STS spectra measured at $T=37$~K on the clean Ta(110) surface and on residual impurities as indicated in (g) using a bare W-tip.  Scanning parameters: (a): $U=500$~mV, $I=0.5$~nA; (b): $U=100$~mV, $I=1$~nA; (c): $U=50$~mV, $I=5$~nA; (d): $U=50$~mV, $I=3$~nA; (g): $U=100$~mV, $I=1$~nA ; (h): $U_{\text{stab}}=1$~V, $I=1$~nA.}
\label{fig:Fig-1}
\end{figure*}

The fresh Ta crystal was initially sputtered for one hour and subsequently annealed at about $1100$~K for 10 min to remove residual Ar and to improve the surface flatness. In this early stage the STM topography of Ta(110) reveals a surface with numerous irregularly shaped step edges, compare Fig.~\ref{fig:Fig-1}(a). In order to identify the main source of contamination AES was performed. Additional peaks are observed at 265~eV and 500~eV, which arise from carbon and oxygen contamination, respectively. After a few cycles of annealing with a total time period of 60 min at temperatures of about 2500~K, the surface morphology is much more flat exhibiting parallel step edges and domains, that vary in their angular orientation, see Fig.~\ref{fig:Fig-1}(b). The observed morphology of Fig.~\ref{fig:Fig-1}(b) resembles characteristic patterns, which have been assigned to an oxygen induced surface reconstruction in case of Nb/NbO-type superstructures~\cite{Hulm1972, Surgers2001, Kuznetsov2010}. The degree of contamination can be reduced if further annealing cycles at 2500~K (total time period of 240 min) are executed, see Fig.~\ref{fig:Fig-1}(c). This high-resolution STM image enables a more detailed analysis of the reconstruction. It forms quasi-ordered structures with $60^\circ$ degrees misaligned domains, consisting of dimer-type chains. Additional annealing cycles not only drastically increase the terrace width, but the surface also exhibits larger areas free of reconstructions. Inside these areas another type of lattice is resolved. The line profiles indicated in Fig.~\ref{fig:Fig-1}(c) are shown in Figs.~\ref{fig:Fig-1}(e,f) and can be used to deduce the lattice constants. The former lattice belonging to the reconstruction reveals a period of $(414 \pm 12)$~pm, while the lattice observed after numerous annealing cycles reveals a period of $(272 \pm 10)$~pm. After the surface quality has been improved further (effective heating time of 25 h at 2500~K), almost the entire surface exhibits the second type of lattice. According to 2D-fast Fourier transform (FFT) analysis, shown in the inset in Fig.~\ref{fig:Fig-1}(d), the surface exhibits a two-fold symmetric lattice with a period of $(285 \pm 13)$~pm. This value corresponds well to the line profile and is in good agreement with the nearest neighbor distance of bare Ta(110)~\cite{Zelikman1973}. Thus, the second lattice is assigned to a clean Ta(110) surface. The angle $\alpha$ between both nearest neighbor directions is estimated to about $\alpha \approx 82^\circ$. This deviates by about $12^\circ$ from the literature value and is probably caused by thermal drift effects, since the STM data were acquired at room temperature.

Finally, a surface almost free of contaminated areas ($>95$~\%) is achieved after applying several subsequent annealing cycles with an overall time period of about 30 h at 2500~K, shown in Fig.~\ref{fig:Fig-1}(g). At this stage, the LEED pattern, displayed in the inset, exhibits clear spots arising from a bcc (110) surface. No additional side spots can be identified. The pattern can be used to cross-check the nearest neighbor distance in comparison to the FFT results. Via LEED a value of $(284 \pm 17)$~pm is found. Further preparation did not substantially decrease the density of the residual defect sites. In addition, all attempts of preparing the surface in a way that the residual O contaminants stick to each other, failed. Thus, they are rather randomly distributed across the entire surface. The resulting Ta(110) surface maintains its purity at a typical background pressure of $p\approx10^{-10}$~mbar for a few tens of hours, only. It is active and easily getters contamination from the residual gas. Fig.~\ref{fig:Fig-1}(h) represents STS spectra of the clean Ta(110) surface. A peak at a bias voltage of about $-500$~mV is found. Note, that this resonance is observed for contaminated areas as well, but with reduced intensity. These observations are in good agreement with results from angle-resolved photoemission experiments~\cite{Kneedler1990}. Based on related Green's function calculations, the surface resonance was identified to originate from Ta's $p_z$ and $d_{yz}$ orbitals~\cite{Hoof1992}.

\subsection{Deposition of Fe on Ta(110)}   

\begin{figure*}[ht!]
\centering
\includegraphics[width=0.75\textwidth]{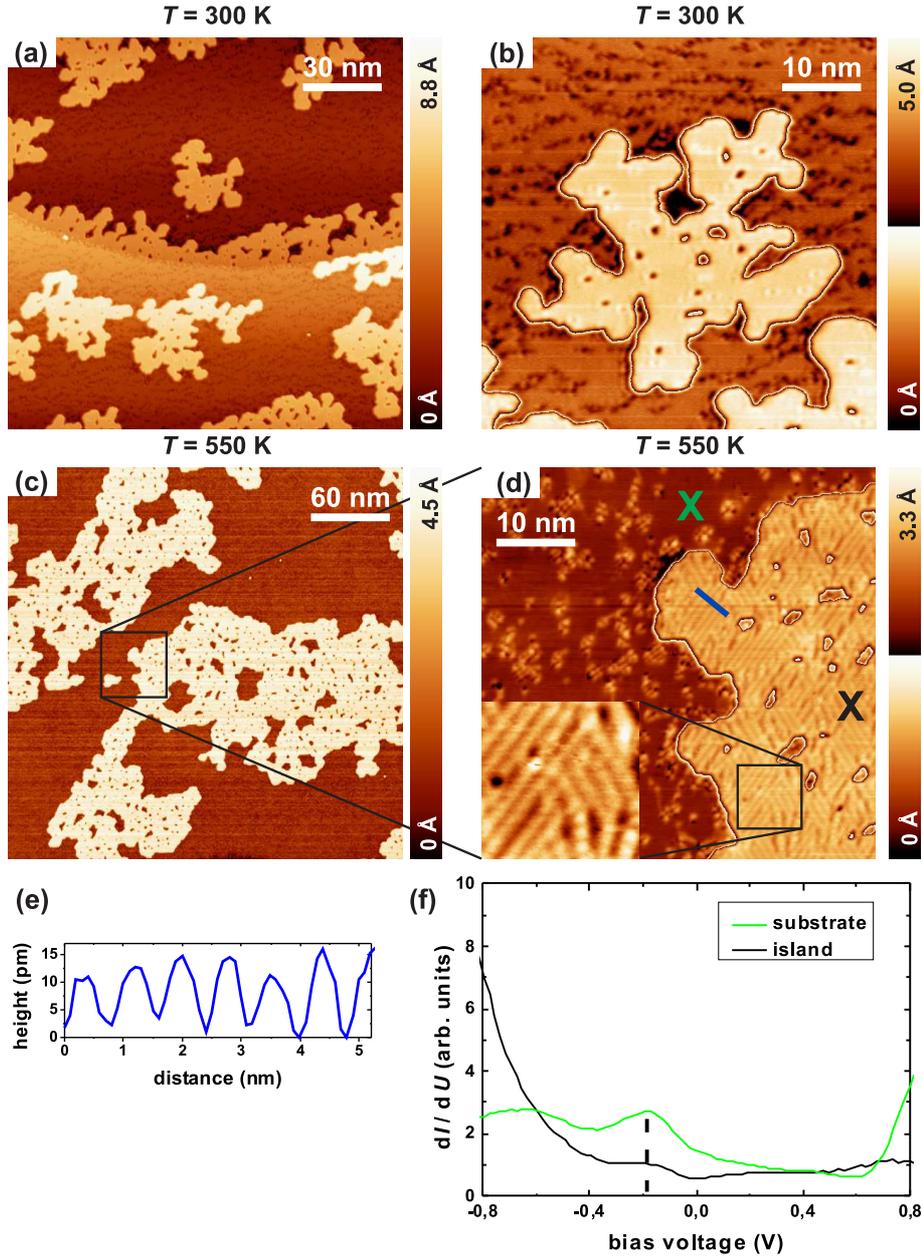}
\caption{Topographies of 0.35~ML Fe deposited on Ta(110) at room temperature with no annealing applied afterward, revealing the surface morphology on a large scale (a) and the typical irregular shape of the Fe islands (b). (c) shows a large scale topography of 0.35~ML Fe/Ta(110) if the sample is subsequently annealed at $T=550$~K for 5~min after deposition. (d) and the according zoom-in show the "network", which is observed on the islands for this preparation. All measurements were performed at $T=37$~K. (e) exhibits a line profile of the "network" as indicated in (d). (f) STS spectra measured by means of a W-tip on the substrate and an island as indicated in (d). Scanning parameters: (a), (b), and (d): $U=100$~mV, $I=0.1$~nA; (c): $U=300$~mV, $I=1$~nA; (f): $U_{\text{stab}}=1$~V, $I=1$~nA.}
\label{fig:Fig-2}
\end{figure*}

Fe was chosen as a model case to study the growth of magnetic nanostructures on Ta(110) since it fulfills the Stoner criterion. A coverage of 0.35 monolayers of Fe was deposited at room temperature. If no annealing is applied, the topography of the sample exhibits two-dimensional growth, see Fig.~\ref{fig:Fig-2}(a). Step-flow grown nanostructures as well as islands appear in an irregular and porous fashion (Fig.~\ref{fig:Fig-2}(b)). We suppose Fe clusters are pinned at randomly distributed residual contaminants of the prepared Ta(110) substrate (see details in Sec. 3.1) during the growth process being the reason for the islands' inhomogeneous shape. It appears as the islands grow around the O contaminants which are still present on the Ta surface. 

In order to increase the surface mobility of the Fe atoms, annealing can be applied during or after the Fe deposition. If the sample is heated at $T = 550$~K for 5 min, the morphology already varies significantly, compare Fig.~\ref{fig:Fig-2}(c). The thermal energy is sufficient so that the Fe islands can coalesce and form larger islands. Although these islands are still porous, their shape is more regular, visible in the Fig.~\ref{fig:Fig-2}(d). In addition, the topography reveals a periodic "network" within the islands, compare the zoom-in in Fig.~\ref{fig:Fig-2}(d). According to the line profile displayed in Fig.~\ref{fig:Fig-2}(e), the "network" exhibits a period of $(800 \pm 20)$~pm. Note, that this period is found only on the islands and not on the substrate terraces. The STS spectra obtained on this sample (Fig.~\ref{fig:Fig-2}(f)) vary strongly from those obtained on bare Ta(110). The surface resonance is shifted in energy towards $-240$~mV indicating a charge transfer between Fe and the Ta substrate. Furthermore, the islands do not reveal any significant spectroscopic feature within the bias range investigated.

\begin{figure*}[ht!]
\centering
\includegraphics[width=0.75\textwidth]{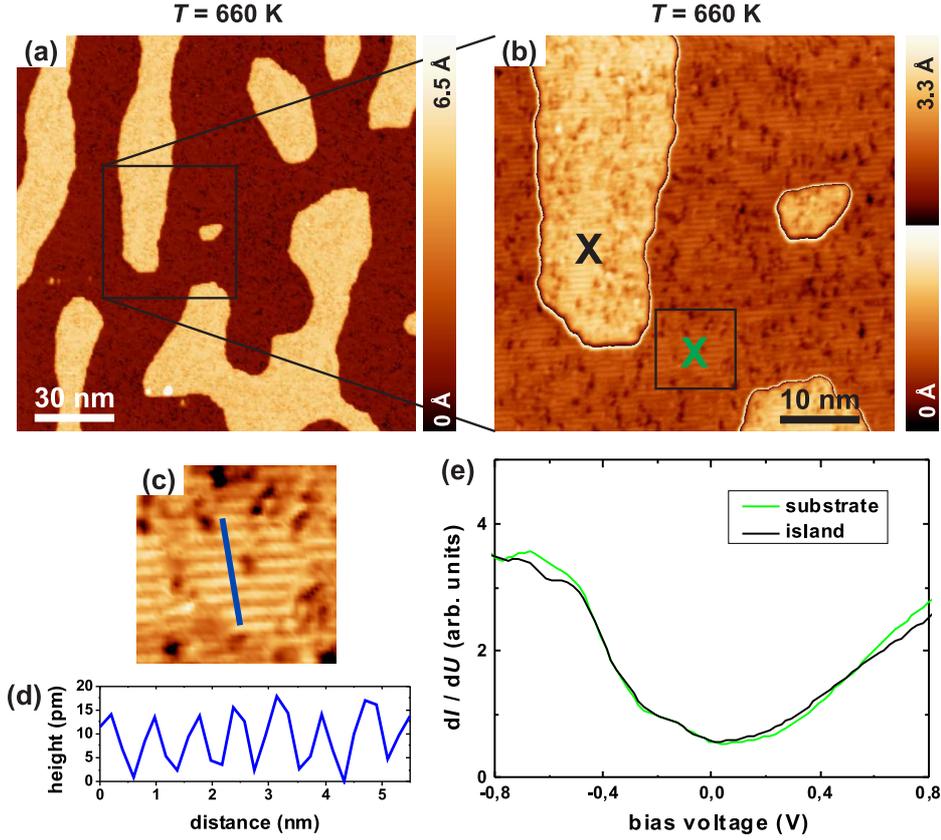}
\caption{(a): STM topography measured at $T=37$~K after room temperature deposition of 0.35~ML Fe on Ta(110) and subsequent annealing to $T=660$~K for 5~min. (b) and (c) are zoom-ins, which reveal a modified surface morphology. The indicated line profile is presented in (d). (e): STS spectra measured by means of a W-tip as indicated in (b). Scanning parameters: (a): $U=100$~mV, $I=1$~nA; (b) $U=-50$~mV, $I=0.5$~nA; (d): $U_{\text{stab}}=1$~V, $I=1$~nA.}
\label{fig:Fig-3}
\end{figure*}

As a next step the Fe film was annealed at $\approx 660$~K to increase the Fe atom's mobility further and to finally form uniform and compact islands. According to Fig.~\ref{fig:Fig-3}(a), the islands appear densely packed and show a regular shape for this type of preparation. Surprisingly, the contrast levels on the islands and the substrate are almost the same and similar stripe patterns are observed on both, islands and substrate, displayed in Fig.~\ref{fig:Fig-3}(b). Note, that the surface morphology does not depend on whether the heating was applied during or after the Fe deposition. According to the zoom-in image of Fig.~\ref{fig:Fig-3}(c) and the indicated line profile (Fig.~\ref{fig:Fig-3}(d)) we find a period of about $(760 \pm 20)$~pm for this preparation on the islands and the substrate. In addition, the STS spectra measured on the substrate and the islands are similar to each other (Fig.~\ref{fig:Fig-3}(e)). The Ta-related resonance cannot be detected any more. These observations hint towards an alloying of Fe and Ta, which has turned the topmost surface layers into a Fe-Ta compound for this type of preparation.

\section{Summary}

The preparation of Ta(110) surfaces studied by means of STM and STS is reported. Extreme solubility and diffusivity of oxygen are the main reasons for the observed superstructures and complicate the preparation of clean Ta(110) surfaces. After multiple annealing cycles at temperatures of about 2500 K under UHV conditions ($\approx 10^{-9}$~mbar), an oxygen depletion zone is formed near the surface. The surface transforms into a predominantly bare Ta(110) surface, almost free of traces of reconstructions ($>95$~\%). By means of STS, a characteristic surface resonance of Ta is found at $-500$~mV in agreement with earlier reports~\cite{Kneedler1990, Hoof1992}.
		
The epitaxial growth of Fe films with coverages below a monolayer reveals different properties of the Fe nanostructures depending on the exact preparation conditions. If no annealing is applied consecutively after the deposition, the Fe islands are irregularly shaped. By annealing at temperatures of about $550$~K the shape of the islands appears to be more compact. However, already at these temperatures a "network" of lines appears within the islands. For annealing temperatures above $660$~K, the morphology of the sample varies even stronger. On the one hand, the islands appear now densely packed, on the other hand, the contrast levels are equal for the substrate and the islands. By STS it is verified that the Ta surface resonance state is no longer preserved, which is explained by an alloying effect at these high annealing temperatures.

\section{Acknowledgements}

Financial support from the ERC Advanced Grant ``ASTONISH'' is gratefully acknowledged.

%\end{multicols}{2}


\section*{References}

\begin{thebibliography}{10}
\expandafter\ifx\csname url\endcsname\relax
  \def\url#1{\texttt{#1}}\fi
\expandafter\ifx\csname urlprefix\endcsname\relax\def\urlprefix{URL }\fi
\expandafter\ifx\csname href\endcsname\relax
  \def\href#1#2{#2} \def\path#1{#1}\fi

\bibitem{Janko1994}
B.~Jank\'{o}, A.~Smith, V.~Ambegaokar, {BCS} superconductivity with fixed
  number parity, Phys. Rev. B 50 (1994) 1152.

\bibitem{Bose2010}
S.~Bose, A.~M. Garc\'{i}a-Garc\'{i}a, M.~M. Ugeda, J.~D. Urbina, C.~H.
  Michaelis, I.~Brihuega, K.~Kern, Observation of shell effects in
  superconducting nanoparticles of {Sn}, Nat. Mater. 9 (2010) 550.

\bibitem{Brihuega2011}
I.~Brihuega, A.~M. Garc\'{i}a-Garc\'{i}a, P.~Ribeiro, M.~M. Ugeda, C.~H.
  Michaelis, S.~Bose, K.~Kern, Experimental observation of thermal fluctuations
  in single superconducting {Pb} nanoparticles through tunneling measurements,
  Phys. Rev. B 84 (2011) 104525.

\bibitem{Binnig1987}
G.~Binnig, H.~Rohrer, Scanning tunneling microscopy -- from birth to
  adolescence, Rev. Mod. Phys. 59 (1987) 615.

\bibitem{Wiesendanger1990}
R.~Wiesendanger, H.-J. G\"{u}ntherodt, G.~G\"{u}ntherodt, R.~J. Gambino,
  R.~Ruf, Observation of vacuum tunneling of spin-polarized electrons with the
  scanning tunneling microscope, Phys. Rev. Lett. 65 (1990) 247.

\bibitem{Suderow2014}
H.~Suderow, I.~Guillamón, J.~G. Rodrigo, S.~Vieira, Imaging superconducting
  vortex cores and lattices with a scanning tunneling microscope, Supercond.
  Sci. Technol. 27 (2014) 063001.

\bibitem{Roditchev2015}
D.~Roditchev, C.~Brun, L.~Serrier-Garcia, J.~C. Cuevas, V.~H.~L. Bessa, M.~V.
  Milošević, F.~Debontridder, V.~Stolyarov, T.~Cren, Direct observation of
  {J}osephson vortex cores, Nat. Phys. 11 (2015) 332.

\bibitem{Kalcheim2012}
Y.~Kalcheim, O.~Millo, M.~Egilmez, J.~W.~A. Robinson, M.~G. Blamire, Evidence
  for anisotropic triplet superconductor order parameter in half-metallic
  ferromagnetic {La}$_{0.7}${Ca}$_{0.3}${Mn}$_{3}${O} proximity coupled to
  superconducting {Pr}$_{1.85}${Ce}$_{0.15}${CuO}$_{4}$, Phys. Rev. B 85 (2012)
  104504.

\bibitem{Ruby2015}
M.~Ruby, B.~W. Heinrich, J.~I. Pascual, K.~J. Franke, Experimental
  demonstration of a two-band superconducting state for lead using scanning
  tunneling spectroscopy, Phys. Rev. Lett. 114 (2015) 157001.

\bibitem{Qin2009}
S.~Qin, J.~Kim, Q.~Niu, C.-K. Shih, Superconductivity at the two-dimensional
  limit, Science 324 (2009) 1314.

\bibitem{Araujo2011}
M.~A.~N. Ara\'{u}jo, A.~M. Garc\'{i}a-Garc\'{i}a, P.~D. Sacramento, Enhancement
  of the critical temperature in iron pnictide superconductors by finite-size
  effects, Phys. Rev. B 84 (2011) 172502.

\bibitem{Brun2014}
C.~Brun, T.~Cren, V.~Cherkez, F.~Debontridder, S.~Pons, D.~Fokin, M.~C.
  Tringides, S.~Bozhko, L.~B. Ioffe, B.~L. Altshuler, D.~Roditchev, Remarkable
  effects of disorder on superconductivity of single atomic layers of lead on
  silicon, Nat. Phys. 10 (2014) 444.

\bibitem{Bose2014}
S.~Bose, P.~Ayyub, A review of finite size effects in quasi-zero dimensional
  superconductors, Rep. Prog. Phys. 77 (2014) 116503.

\bibitem{Linder2015}
J.~Linder, J.~W.~A. Robinson, Superconducting spintronics, Nat. Phys. 11 (2015)
  307.

\bibitem{Nadj2014}
S.~Nadj-Perge, I.~K. Drozdov, J.~Li, H.~Chen, S.~Jeon, J.~Seo, A.~H.
  {MacDonald}, B.~A. Bernevig, A.~Yazdani, Observation of majorana fermions in
  ferromagnetic atomic chains on a superconductor, Science 346 (2014) 602.

\bibitem{Wortelen2015}
H.~Wortelen, K.~Miyamoto, H.~Mirhosseini, T.~Okuda, A.~Kimura, D.~Thonig,
  J.~Henk, M.~Donath, Spin-orbit influence on $d_{z^2}$-type surface state at
  \normalsize{Ta(110)}, Phys. Rev. B 92 (2015) 161408(R).

\bibitem{Sewell1972}
P.~B. Sewell, D.~F. Mitchell, M.~Cohen, A kinetic study of the initial
  oxidation of a \normalsize{Ta}(110) surface using oxygen $k_{\alpha}$ x-ray
  emission, Surf. Sci. 29 (1972) 173.

\bibitem{Huger2005}
E.~H{\"u}ger, H.~Wormeester, K.~Osuch, Subsurface miscibility of metal
  overlayers with \normalsize{V}, \normalsize{Nb} and \normalsize{Ta}
  substrates, Surf. Sci. 580 (2005) 173.

\bibitem{Garg1996}
S.~P. Garg, N.~Krishnamurthy, A.~A. amd M.~Venkatraman, The \normalsize{O-Ta
  (Oxygen-Tantalum)} system, J. Phase Equilib. 17 (1996) 63.

\bibitem{Hulm1972}
J.~K. Hulm, C.~K. Jones, R.~A. Hein, J.~W. Gibson, Superconductivity in the
  \normalsize{TiO and NbO} systems, J. Low Temp. Phys. 7 (1972) 291.

\bibitem{Surgers2001}
C.~S{\"u}rgers, M.~Sch{\"o}ck, H.~von L{\"o}hneysen, Oxygen-induced surface
  structure of \normalsize{Nb(110)}, Surf. Sci. 471 (2001) 209.

\bibitem{Kuznetsov2010}
M.~V. Kuznetsov, A.~S. Razinkin, A.~L. Ivanovskii, Oxide nanostructures on a
  \normalsize{Nb} surface and related systems: experiments and \textit{ab
  initio} calculations, Phys. Usp. 53 (2010) 995.

\bibitem{Zelikman1973}
A.~N. Zelikman, G.~A. Meerson, Metallurgy of rare metals, Metalurgya, Moscow
  1973.

\bibitem{Kneedler1990}
E.~Kneedler, D.~Skelton, K.~E. Smith, S.~D. Kevan,
  Surface-state-surface-resonance transition on {Ta}(011), Phys. Rev. Lett. 64
  (1990) 3151.

\bibitem{Hoof1992}
J.~B.~A.~N. van Hoof, S.~Crampin, J.~E. Inglesfield, The surface state-surface
  resonance transition on {Ta}(011), J. Phys.: Condens. Matter 4 (1992) 8477.

\end{thebibliography}
\end{document}